%% file: paper.tex
\documentclass[twocolumn,showpacs,aps,prl,superscriptaddress,xspace]{revtex4}

\usepackage{graphicx}
\usepackage{dcolumn}
\usepackage{amsmath}
\usepackage{amssymb}
\usepackage{epsfig}

\input pubboard/babarsym

\newcommand{\BABARPubYear}    {07}
\newcommand{\BABARPubNumber}  {062}
\newcommand{\SLACPubNumber} {12807}

\newcommand{\mmiss}{\ensuremath{m_\mathrm{miss}^2}\xspace}
\newcommand{\ds}{\ensuremath{D^{(*)}}\xspace}
\newcommand{\eextra}{\ensuremath{E_\mathrm{extra}}\xspace}
\newcommand{\btag}{\ensuremath{B_\mathrm{tag}}\xspace}
\newcommand{\pstarl}{\ensuremath{|{\bf p}^*_\ell|}\xspace}
\newcommand{\gevccnosp}{\ensuremath{{\mathrm{Ge\kern -0.1em V\!/}c^2}}\xspace}

\begin{document}

\preprint{\babar-PUB-\BABARPubYear/\BABARPubNumber} 
\preprint{SLAC-PUB-\SLACPubNumber} 

\begin{flushleft}
%\babar-CONF-\BABARPubYear/\BABARPubNumber\\
%SLAC-PUB-\SLACPubNumber\\
%hep-ex/\LANLNumber\\[10mm]
\end{flushleft}

\title{
{\large \bf
\boldmath Observation of the Semileptonic Decays $B\to\Dstar\taum\nutb$ and Evidence for $B\to D\taum\nutb$}
}

\input pubboard/authors_aug2007.tex

\date{\today}

\begin{abstract}
We present measurements of the semileptonic decays 
$\Bm\to\Dz\taum\nutb$, $\Bm\to\Dstarz\taum\nutb$,
$\Bzb\to\Dp\taum\nutb$, and $\Bzb\to\Dstarp\taum\nutb$,
which are potentially sensitive to non--Standard Model amplitudes.
The data sample comprises $232\times 10^6\ \FourS\to\BB$ decays
collected with the \babar\ detector. 
From a combined fit to \Bm and \Bzb channels, we obtain the branching fractions
${\cal B}(B\to D\taum\nutb)=(0.86\pm 0.24\pm 0.11\pm 0.06)\%$ 
and ${\cal B}(B\to D^*\taum\nutb)=(1.62\pm 0.31\pm 0.10\pm 0.05)\%$  
(normalized for the \Bzb),
where the uncertainties are statistical, systematic, and normalization-mode-related.
\vspace{3mm}

\end{abstract}


\maketitle

%%% Introduction
Semileptonic decays of $B$ mesons to the $\tau$ 
lepton---the heaviest of the three charged leptons---provide 
a new source of 
information on Standard Model (SM) 
processes~\cite{KornerSchuler,Falk_etal,HuangAndKim}, 
as well as a new window on physics beyond 
the SM~\cite{GrzadkowskiAndHou,Tanaka,KiersAndSoni,Itoh,ChenAndGeng}.
In the SM, semileptonic decays occur at tree level and are mediated by the $W$ boson, but
the large mass of the $\tau$ lepton provides sensitivity to 
additional amplitudes, such as those mediated by a charged Higgs 
boson. Experimentally, $b\to c\taum\nutb$ decays~\cite{ChargeConjugate} are 
challenging because the
final state contains not just one, but two or three neutrinos as a
result of the $\tau$ decay.

%% Paragraph on theory predictions
Branching fractions for semileptonic $B$ decays to $\tau$ leptons are predicted to be
smaller than those for $\ell=e,\ \mu$~\cite{Ell}.
Calculations based on the SM predict
${\cal B}(\Bzb\to\Dp\taum\nutb)=(0.69\pm0.04)\%$ and
${\cal B}(\Bzb\to\Dstarp\taum\nutb)=(1.41\pm0.07)\%$~\cite{ChenAndGeng},
which account for most of the predicted inclusive rate
${\cal B}(B\to X_c\taum\nutb)=(2.30\pm0.25)\%$~\cite{Falk_etal}
(here, $X_c$ represents all
hadronic final states from the $b\to c$ transition).
Calculations~\cite{GrzadkowskiAndHou,Tanaka,KiersAndSoni,Itoh,ChenAndGeng} 
in multi-Higgs doublet models show that substantial
departures, either positive or negative, from the SM decay rate could occur for
${\cal B}(B\to D\taum\nutb)$. Those for  
${\cal B}(B\to\Dstar\taum\nutb)$, however, are expected to be smaller. 

Theoretical predictions for semileptonic decays to exclusive final states 
require knowledge of the form factors, which parametrize the hadronic current
as functions of $q^2=(p_B-p_{\ds})^2.$ For light leptons ($e$, $\mu$), there is effectively
one form factor for $B\to D\ellm\nulb$, while there are 
three for $B\to\Dstar\ellm\nulb$. If a $\tau$ lepton is produced instead,
one additional form factor enters in each mode. The form factors for $B\to\ds\ellm\nulb$ 
decays involving the light leptons have been measured~\cite{FF}. 
Heavy quark symmetry (HQS) relations~\cite{IsgurWise} allow one to express
the two additional form factors for $B\to\ds\taum\nutb$ in terms of the
form factors measurable from decays with the light leptons. With sufficient data, one could
probe the additional form factors and test the HQS relations.

%% Paragraph on previous meaurements
The first measurements of semileptonic $b$-hadron decays to $\tau$ leptons were performed by 
the LEP experiments~\cite{LEP} operating at the \Z resonance,
yielding an average~\cite{PDG} inclusive branching fraction
${\cal B}(b_{\rm had}\to X\taum\nutb)=(2.48\pm 0.26)\%$, where $b_{\rm had}$ 
represents the mixture of
$b$-hadrons produced in $\Z\to\bbbar$ decays. The Belle experiment has
recently obtained ${\cal B}(\Bzb\to\Dstarp\taum\nutb)=(2.02^{+0.40}_{-0.37}\pm0.37)\%$ 
\cite{Belle}.

We determine the branching fractions of
four exclusive decay modes: $\Bm\to\Dz\taum\nutb$,
$\Bm\to\Dstarz\taum\nutb$, $\Bzb\to\Dp\taum\nutb$, and $\Bzb\to\Dstarp\taum\nutb$,
each of which is measured relative to the corresponding $e$ and $\mu$
modes. To reconstruct the $\tau$, we use the decays
$\taum\to\en\nueb\nut$ and $\taum\to\mun\numb\nut$, which are experimentally most
accessible. The main challenge of the
measurement is to separate $B\to\ds\taum\nutb$ decays, which have
three neutrinos, from $B\to\ds\ellm\nulb$ decays, which have
the same observable final-state particles but only one neutrino.

%% Paragraph on data sample and BaBar detector
We analyze data collected with the \babar\ detector~\cite{BABARNIM}
at the \pep2 \epem storage rings at 
the Stanford Linear Accelerator Center. The data sample used
comprises $208.9\ \invfb$ of integrated luminosity recorded on the $\FourS$
resonance, yielding $232\times 10^6\ \BB$ decays.

%%% Analysis overview and strategy

The analysis strategy is to reconstruct the decays of both $B$ mesons in the $\FourS\to\BB$ 
event, providing powerful constraints on unobserved particles. One $B$ 
meson, denoted \btag, is fully reconstructed in a purely hadronic decay chain. The remaining
charged particles and photons are required to be consistent with the products of a $b\to c$
semileptonic $B$ decay: a hadronic system, a \ds meson,
and a lepton ($e$ or $\mu$). The lepton may be either primary or from
$\taum\to\ellm\nulb\nut$.
We calculate the missing four-momentum 
$p_\mathrm{miss}=[p_{\epem}-p_{\rm tag}-p_{\ds}-p_{\ell}]$ 
of any particles recoiling against the 
observed $\btag+\ds\ell$ system. A large peak at zero in $\mmiss= p_\mathrm{miss}^2$ corresponds 
to semileptonic decays with one neutrino, whereas signal events form a broad tail out to 
$\mmiss\sim 8\ (\gevccnosp)^2$. To separate signal and background events, we 
perform a fit to the joint distribution of \mmiss and the lepton 
momentum (\pstarl) in the rest frame of the $B$ meson.
In signal events, the observed lepton is the daughter of the $\tau$ 
and typically has a soft spectrum; for most background events, this lepton 
typically has higher momentum.

%%% Event selection
We reconstruct \btag candidates~\cite{BtagAlgorithm} in 1114 final states
$\btag\to\ds Y^\pm$.
Tag-side \ds candidates are reconstructed in 21 decay chains, and the
$Y^\pm$ system may consist
of up to six light hadrons (\pipm, \piz, \Kpm, or \KS). \btag candidates
are identified using two kinematic variables, $\mes=\sqrt{s/4-|{\bf p}_\mathrm{tag}|^2}$
and $\DeltaE=E_\mathrm{tag}-\sqrt{s}/2$, where $\sqrt{s}$ is the total \epem
energy, $|{\bf p}_\mathrm{tag}|$ is the magnitude of the \btag momentum, and $E_\mathrm{tag}$ is
the \btag energy, all defined in the \epem center-of-mass frame. We require
$\mes>5.27\ \gevcc$ and $|\DeltaE|<72\ \mev$, corresponding
to $\pm 4\sigma$ (standard deviations). We reconstruct \btag candidates
in approximately $0.3\%$ to $0.5\%$ of \BB events.

For the $B$ meson decaying semileptonically, we reconstruct \ds candidates in the modes $\Dz\to\Km\pip$,
$\Km\pip\piz$, $\Km\pip\pipi$, $\KS\pipi$; $\Dp\to\Km\pip\pip$,
$\Km\pip\pip\piz$, $\KS\pip$, $\Km\Kp\pip$; $\Dstarz\to\Dz\piz$, $\Dz\gamma$;
and $\Dstarp\to\Dz\pip$, $\Dp\piz$. $D$ (\Dstar) candidates are selected within
$4\sigma$ of the $D$ mass ($\Dstar-D$ mass difference), with $\sigma$
typically $5$--$10\ \mevcc$ ($1$--$2\ \mevcc$).
Electron candidates must have lab-frame momentum $|{\bf p}_e|>300 \mevc$;
muon candidates must have an appropriate 
signature in the muon detector system, effectively requiring  
$|{\bf p}_{\mu}|\gtrapprox 600 \mevc$. 
The energy of electron candidates is corrected for bremsstrahlung energy loss
if photons are found close to the electron direction. 

We require that all charged tracks be associated with either
the \btag, \ds, or $\ell$ candidate. We compute \eextra, the sum of the
energies of all photon candidates not associated with the $\btag+\ds\ell$ 
candidate system, and we require $\eextra<150$--$300\ \mev$, depending
on the \ds channel. We suppress hadronic events and combinatorial
backgrounds by requiring $|{\bf p}_\mathrm{miss}|>200\ \mevc$ and $q^2>4\ (\gevccnosp)^2$.
If multiple candidate systems pass this selection, we select the
one with the lowest value of \eextra. To improve the \mmiss resolution,
we perform a kinematic fit to the event, constraining particle masses
to known values and requiring tracks from $B$, $D$, and \KS
mesons to originate from appropriate common vertices.
All event selection requirements and fit procedures
have been defined using simulated events or using control samples
in data that exclude the signal region.

\begin{figure}[tb!]
\includegraphics[width=3.1in]{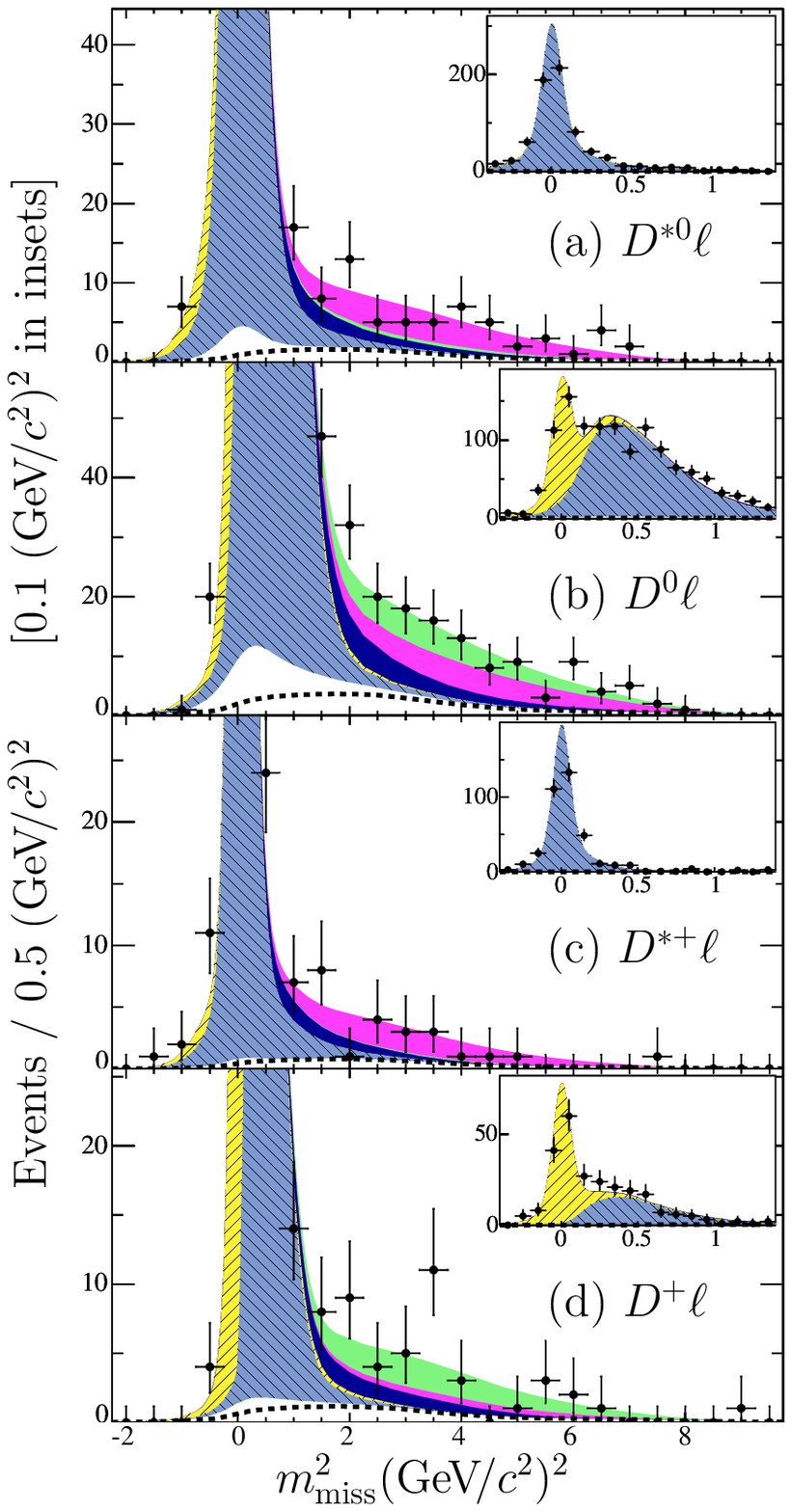}
\caption {(Color online) Distributions of events and fit projections in \mmiss 
for the four final states:
$\Dstarz\ell^-$, $\Dz\ell^-$, $\Dstarp\ell^-$, and $\Dp\ell^-$. 
The normalization region $\mmiss\approx 0$ is shown with finer binning in the insets. 
The fit components are combinatorial background (white, below dashed line), 
charge crossfeed background (white, above dashed line), the $B\to D\ellm\nulb$ 
normalization mode (// hatching, yellow), the $B\to\Dstar\ellm\nulb$ normalization
mode ($\backslash\backslash$ hatching, light blue), $B\to D^{**}\ellm\nulb$ background (dark, or blue),
the $B\to D\taum\nutb$ signal (light grey, green), and the $B\to\Dstar\taum\nutb$ signal
(medium grey, magenta). The fit shown incorporates the \Bm--\Bzb constraints.
}\label{fig:fit}
\end{figure}

Figure~\ref{fig:fit} shows the distributions of \mmiss for the
four $\ds\ell$ channels, along with the projections 
of the maximum likelihood fit discussed below. 
We observe large peaks at $\mmiss \approx 0$ as well as
events in the signal region at large \mmiss.
The peaks 
are mainly due to $B\to\ds\ellm\nulb$, which serve as normalization modes.
The structure of this 
background is shown in the inset figures, which expand the region 
$-0.4<\mmiss< 1.4\ (\gevccnosp)^2$.
$B\to\Dstar\ellm\nulb$ background is the dominant feature in
the two $\Dstar\ell$ channels (Figs.~\ref{fig:fit}a, c);
the two $D\ell$ channels (Figs.~\ref{fig:fit}b, d) are dominated
by $B\to D\ellm\nulb$ decays but also include substantial
contributions from true \Dstar mesons where the low-momentum \piz
or photon from $\Dstar\to D\piz$ or $D\gamma$ is not reconstructed.
Similarly, $B\to\Dstar\taum\nutb$ events can feed down to the
$D\ell$ channels. The fit therefore includes feed-down
components for both the signal and normalization modes,
as well as smaller feed-up contributions from
$B\to D(\ellm/\taum)\nub$ into the $\Dstar\ell$ channels.
Other sources of background include
$B\to D^{**}(\ellm/\taum)\nub$ events (here $D^{**}$
represents charm resonances heavier than the $\Dstar(2010)$, as well as
non-resonant $\ds n\pi$ systems with $n\geq 1$); charge-crossfeed 
($B\to\ds\ellm\nulb$ events reconstructed with the wrong charge for the
\btag and \ds meson, typically because a low-momentum $\pipm$ is swapped between
them); and combinatorial background. This last background is dominated by
hadronic $B$ decays, such as $B\to\ds D_s^{(*)}$, that produce a secondary lepton,
including $\tau$ leptons from $D_s$ decay.

To constrain $B\to D^{**}(\ellm/\taum)\nub$ background, we select
four control samples, identical to the signal channels but 
in which an extra \piz meson is observed. Most of the $D^{**}$ background in the
signal channels occurs when the \piz from $D^{**}\to\ds\piz$ is not
reconstructed, so these control samples provide a normalization of
the background source. $D^{**}$ decays in which a $\pipm$ is lost do not have the correct
charge correlation between the \btag and \ds, and decays with two missing charged
pions are rare. The feed-down probabilities for the $D^{**}(\ellm/\taum)\nub$
background are determined from simulation, with uncertainties in the
$D^{**}$ content treated as a systematic error. 

%%% The fit
We perform a relative measurement, extracting both signal $B\to\ds\taum\nutb$
and normalization $B\to\ds\ellm\nulb$ yields from the fit to obtain the 
four branching ratios $R(\Dz)$, $R(\Dp)$, $R(\Dstarz)$, and $R(\Dstarp)$, where,
for example, $R(\Dstarz)\equiv{\cal B}(\Bm\to\Dstarz\taum\nutb)/{\cal B}(\Bm\to\Dstarz\ellm\nulb)$. 
These ratios are normalized such that $\ell$ represents only one of $e$ or $\mu$;
however, both light lepton species are included in the measurement.
Signal and background yields are extracted 
using an extended, unbinned maximum likelihood fit to the 
joint (\mmiss, \pstarl) distribution. The 18-parameter fit
is performed simultaneously in the four signal channels and the four $D^{**}$
control samples. In each of the four
signal channels, we describe the data as the sum of seven components (shown in Fig.~\ref{fig:fit}):
$D\taum\nutb$, $\Dstar\taum\nutb$, $D\ellm\nulb$, $\Dstar\ellm\nulb$, $D^{**}(\ellm/\taum)\nub$,
charge crossfeed, and combinatorial background. The four $D^{**}$ control samples
are described as the sum of five components: $D^{**}(\ellm/\taum)\nub$, $D\ellm\nulb$,
$\Dstar\ellm\nulb$, charge crossfeed, and combinatorial background. 
Probability distribution functions (PDFs) are primarily
determined from simulated event samples.
Both the signal and normalization modes
are described using HQET-based form factors~\cite{CLN} for
which the parameters and their uncertainties are determined by 
experimental measurements~\cite{FF}. Parameters describing
the amount of the dominant feed-down components---$\Dstar$ feed-down into 
the $D$ channels---are determined directly by the fit.

%%% Results
Table~\ref{tab:results} summarizes the results from two fits, one in
which all four signal yields can vary independently, and a second
fit in which we constrain~\cite{Isospin} 
$R(\Dp)=R(\Dz)$ and $R(\Dstarp)=R(\Dstarz)$. The \mmiss projections
shown in Fig.~\ref{fig:fit} are those from this \Bm--\Bzb-constrained fit.

\begin{table*}
\caption{Results from fits to data:
the signal yield ($N_\mathrm{sig}$), the yield of normalization $B\to\ds\ellm\nulb$ events ($N_\mathrm{norm}$),
the ratio of signal and normalization mode efficiencies ($\varepsilon_\mathrm{sig}/\varepsilon_\mathrm{norm}$),
the relative systematic error due to the fit yields ($(\Delta R/R)_\mathrm{fit}$), the
relative systematic error due to the efficiency ratios ($(\Delta R/R)_\varepsilon$),
the branching fraction relative to the normalization mode ($R$), the absolute
branching fraction ($\cal B$), and the total and statistical signal significances
($\sigma_\mathrm{tot}$ and $\sigma_\mathrm{stat}$).
The first two errors on $R$
and $\cal B$ are statistical and systematic, respectively; the third error on $\cal B$
represents the uncertainty on the normalization mode~\cite{NormBF}.
The last two rows show the results of the fit with the \Bm--\Bzb constraint applied,
where $\cal B$ is expressed for the \Bzb. The statistical correlation between
$R(D)$ and $R(\Dstar)$ in this fit is $-0.51$.
}
\label{tab:results}
\begin{tabular}{l l r@{$\pm$}l r@{$\pm$}l c c c r@{$\pm$}r@{$\pm$}r l@{$\pm$}l@{$\pm$}l@{$\pm$}l l}\\ \hline\hline
\multicolumn{2}{l}{Mode} & \multicolumn{2}{c}{$N_\mathrm{sig}$} &
  \multicolumn{2}{c}{$N_\mathrm{norm}$} &
  \multicolumn{1}{c}{$\varepsilon_\mathrm{sig}/\varepsilon_\mathrm{norm}$} &
  \multicolumn{1}{c}{$(\Delta R/R)_\mathrm{fit}$} &
  \multicolumn{1}{c}{$(\Delta R/R)_\varepsilon$} &
  \multicolumn{3}{c}{$R$} &
  \multicolumn{4}{c}{$\cal B$} & \multicolumn{1}{c}{$\sigma_\mathrm{tot}$} \\
\multicolumn{6}{c}{} &
  \multicolumn{1}{c}{} & \multicolumn{1}{c}{[\%]} &
  \multicolumn{1}{c}{[\%]} &
  \multicolumn{3}{c}{[\%]} & \multicolumn{4}{c}{[\%]} &
  \multicolumn{1}{c}{$(\sigma_\mathrm{stat})$} \\ \hline
\Bm & $\to\Dz\taum\nutb$ & $35.6$ & $ 19.4$ & $347.9$ & $23.1$ & $1.85$ & $15.5$ & $1.6$ &
  $31.4$ & $17.0$ & $4.9$ & $0.67$ & $0.37$ & $0.11$ & $0.07$ & $1.8\ (1.8)$ \\
\Bm & $\to\Dstarz\taum\nutb$ & $92.2$ & $19.6$ & $1629.9$ & $63.6$ & $0.99$ & $9.7$ & $1.5$ &
  $34.6$ & $7.3$ & $3.4$ & $2.25$ & $0.48$ & $0.22$ & $0.17$ & $5.3\ (5.8)$ \\
\Bzb & $\to\Dp\taum\nutb$ & $23.3$ & $7.8$ & $150.2$ & $13.3$ & $1.83$ & $13.9$ & $1.8$ &
  $48.9$ & $16.5$ & $6.9$ & $1.04$ & $0.35$ & $0.15$ & $0.10$ & $3.3\ (3.6)$ \\
\Bzb & $\to\Dstarp\taum\nutb$ & $15.5$ & $7.2$ & $482.3$ & $25.5$ & $0.91$ & $3.6$ & $1.4$ &
  $20.7$ & $9.5$ & $0.8$ & $1.11$ & $0.51$ & $0.04$ & $0.04$ & $2.7\ (2.7)$ \\ \hline
$B$ & $\to D\taum\nutb$ & $66.9$ & $18.9$ & $497.8$ & $26.4$ & --- & $12.4$ & $1.4$ &
  $41.6$ & $11.7$ & $5.2$ & $0.86$ & $0.24$ & $0.11$ & $0.06$ & $3.6\ (4.0)$ \\
$B$ & $\to\Dstar\taum\nutb$ & $101.4$ & $19.1$ & $2111.5$ & $ 68.1$ & --- & $5.8$ & $1.3$ &
  $29.7$ & $5.6$ & $1.8$ & $1.62$ & $0.31$ & $0.10$ & $0.05$ & $6.2\ (6.5)$ \\ \hline\hline
\end{tabular}
\end{table*}

%%% Checks
The features of the event sample have been extensively checked.  
The observed lepton spectra are well described by the fit 
both in signal- and in background-dominated
regions. The properties of reconstructed \btag mesons,
such as charged and neutral 
daughter multiplicities, are consistent with expectations. 
Control samples of $B\to\ds\ellm\nulb$
events, kinematically selected without a cut on $\mmiss$, provide
checks of numerous distributions, including the $\mmiss$ tails.
 
%%% Systematics
Systematic uncertainties on $R$ associated with the fit,
$(\Delta R/R)_\mathrm{fit}$ in Table~\ref{tab:results},
are determined by running ensembles of fits in which
input parameters are distributed according to our knowledge of the
underlying source, and include the PDF parametrization
(2\% to 12\%); the composition of combinatorial
backgrounds (2\% to 11\%); the mixture of $D^{**}$ states in
$B\to D^{**}\ellm\nulb$ decays ($0.3$\% to 6\%); the $B\to\Dstar$
form factors (0.2\% to 1.9\%); the \piz reconstruction efficiency, which affects the
$\Dstar$ and $D^{**}$ feed-down rates (0.5\% to 1.1\%);
and the \mmiss resolution for $B\to\Dstar\ellm\nulb$ events
($0.1$\% to $1.6$\%). Uncertainties on the
$B\to D$ form factors contribute less than 1\%.
Uncertainties on $R$ propagated from the ratio of efficiencies,
$(\Delta R/R)_{\varepsilon}$ in Table~\ref{tab:results},
are typically small due to cancellations,
and include the limited statistics in the simulation ($0.8\%$ to $1.5\%$)
and systematic errors related to detector performance ($0.2\%$ to $0.7\%$).
Uncertainties from modeling
final-state radiation are $0.3$\% to $0.5$\%; uncertainties on
the branching fractions of the reconstructed modes contribute $0.3$\% or less.
Finally, the uncertainty on 
$\cal B(\taum\to\ellm\nulb\nut)$~\cite{PDG} contributes $0.2\%$ to all modes.

Table~\ref{tab:results} gives the significances of the signal yields. 
The statistical significance is determined from $\sqrt{2\Delta(\ln\cal L)}$,
where $\Delta(\ln\cal L)$ is the change in log-likelihood between the nominal fit and
the no-signal hypothesis. The total significance is determined by
including $(\Delta R/R)_\mathrm{fit}$ in quadrature with the statistical error.

%%% Conclusions
We have presented measurements of the decays
$B\to D\taum\nutb$ and $B\to\Dstar\taum\nutb$, relative to the
corresponding decays to light leptons. We find
$R(D)=(41.6\pm 11.7\pm 5.2)\%$ and
$R(\Dstar)=(29.7\pm 5.6\pm 1.8)\%$,
where the first error is statistical and the second is systematic.
Normalizing to known \Bzb branching fractions~\cite{NormBF}, we obtain
\begin{eqnarray}
{\cal B}(B\to D\taum\nutb)&=&(0.86\pm 0.24\pm 0.11\pm 0.06)\% \nonumber\\
{\cal B}(B\to \Dstar\taum\nutb)&=&(1.62\pm 0.31\pm 0.10\pm 0.05)\%,\nonumber
\end{eqnarray}
\noindent where the third error is from that on the normalization
mode branching fraction.
The significances of the signals are $3.6\sigma$ and $6.2\sigma$, respectively.
The modes $\Bm\to\Dz\taum\nutb$, $\Bm\to\Dstarz\taum\nutb$, and
$\Bzb\to\Dp\taum\nutb$ have not been studied previously,
while the measurement of $\Bzb\to\Dstarp\taum\nutb$ 
is consistent with the Belle result~\cite{Belle}.
The averaged branching fractions 
are about $1\sigma$ higher than the SM predictions but, given the
uncertainties, there is still room for a sizeable non-SM contribution. 

%%% Input the pubboard acknowledgements file
\input pubboard/acknow_PRL.tex

\end{document}

%% file: pubboard/authors_aug2007.tex
%% author list as of 01-Aug-2007 (565 authors)
%
\author{B.~Aubert}
\author{M.~Bona}
\author{D.~Boutigny}
\author{Y.~Karyotakis}
\author{J.~P.~Lees}
\author{V.~Poireau}
\author{X.~Prudent}
\author{V.~Tisserand}
\author{A.~Zghiche}
\affiliation{Laboratoire de Physique des Particules, IN2P3/CNRS et Universit\'e de Savoie, F-74941 Annecy-Le-Vieux, France }
\author{J.~Garra~Tico}
\author{E.~Grauges}
\affiliation{Universitat de Barcelona, Facultat de Fisica, Departament ECM, E-08028 Barcelona, Spain }
% ...
\collaboration{The \babar\ Collaboration}
\noaffiliation

%% file: pubboard/acknow_PRL.tex
We are grateful for the excellent luminosity and machine conditions
provided by our \pep2\ colleagues, 
and for the substantial dedicated effort from
the computing organizations that support \babar.
The collaborating institutions wish to thank 
SLAC for its support and kind hospitality. 
This work is supported by
DOE
and NSF (USA),
NSERC (Canada),
CEA and
CNRS-IN2P3
(France),
BMBF and DFG
(Germany),
INFN (Italy),
FOM (The Netherlands),
NFR (Norway),
MIST (Russia),
MEC (Spain), and
STFC (United Kingdom). 
Individuals have received support from the
Marie Curie EIF (European Union) and
the A.~P.~Sloan Foundation.

%% file: paper.bbl
\begin{thebibliography}{99}

% SM predictions 
\bibitem{KornerSchuler} J.G.~K\"orner and G.A.~Schuler, Phys.~Lett.~B {\bf 231}, 306 (1989);
Z.~Phys.~C {\bf 46}, 93 (1990).
\bibitem{Falk_etal} A.F.~Falk {\it et al.}, Phys.~Lett.~B {\bf 326}, 145 (1994).
\bibitem{HuangAndKim} D.~S.~Hwang, and D.-W.~Kim, \epjc{14}, 271 (2000). 

% new physics predictions
\bibitem{GrzadkowskiAndHou} B.~Grz\c{a}dkowski and W.-S.~Hou, Phys.~Lett.~B {\bf 283}, 427 (1992).
\bibitem{Tanaka} M.~Tanaka, Z.~Phys.~C {\bf 67}, 321 (1995).
\bibitem{KiersAndSoni} K.~Kiers and A.~Soni, \jprd{56}, 5786 (1997).
\bibitem{Itoh} H.~Itoh, S.~Komine, and Y.~Okada, Prog. Theor. Phys. {\bf 114} 179 (2005).


\bibitem{ChenAndGeng} C.-H.~Chen and C.-Q.~Geng, JHEP {\bf 0610}, 053 (2006).
\bibitem{ChargeConjugate} Charge-conjugate modes are implied throughout.

\bibitem{Ell}Throughout this letter, we use the symbol $\ell$ to refer only to the light charged
  leptons $e$ and $\mu$. The symbol \ds refers either to a $D$ or a \Dstar meson.
\bibitem{FF} CLEO Collab., J.E.~Duboscq {\it et al.}, Phys.~Rev.~Lett.~{\bf 76}, 3898 (1996);
  \babar\ Collab., B.~Aubert {\it et al.}, Phys.~Rev.~D {\bf 74}, 092004 (2006);
  HFAG, E.~Barberio {\it et al.}, arXiv:0704.3575.
\bibitem{IsgurWise} N.~Isgur and M.B.~Wise, Phys.~Lett.~B {\bf 232}, 113 (1989); B {\bf 237},
527 (1990).

\bibitem{LEP} L3 Collab., M.~Acciarri {\it et al.}, \plb{332}, 201 (1994);
  \zpc{71}, 379 (1996);
ALEPH Collab., R.~Barate {\it et al.}, \epjc{6}, 555 (2001);
OPAL Collab., G.~Abbiendi {\it et al.}, \plb{520}, 1 (2001).
\bibitem{PDG} Particle Data Group, W.-M.~Yao {\it et al.}, J.~Phys.~G: Nucl.~Part.~Phys.~{\bf 33}, 1 (2006).
\bibitem{Belle} Belle Collab., A. Matyja {\it et al.}, arXiv:0706.4429, submitted to Phys.~Rev.~Lett.
\bibitem{BABARNIM} \babar\ Collab., B.~Aubert {\it et al.}, \nima{479}, 1 (2002).
\bibitem{BtagAlgorithm} \babar\ Collab., B.~Aubert {\it et al.}, \jprl{92}, 071802 (2004).
\bibitem{CLN} I.~Caprini, L.~Lellouch, and M.~Neubert, \npb{530} 153 (1998).
\bibitem{Isospin} This constraint follows from isospin symmetry in both the signal
and normalization modes but is more general.
\bibitem{NormBF} We use~\cite{PDG} to normalize the four individual branching fractions. For
the \Bm--\Bzb-constrained measurement, we use our own averages of the values in~\cite{PDG}:
${\cal B}(\Bzb\to\Dp\ellm\nulb)=(2.07\pm 0.14)\%$ and
${\cal B}(\Bzb\to\Dstarp\ellm\nulb)=(5.46\pm 0.18)\%$.
\end{thebibliography}
